\documentclass[showkeys,nofootinbib,pra,twocolumn]{revtex4}
\usepackage{graphicx}

\begin{document}

\title{Wholeness as a Conceptual Foundation of Physical Theories}
\date{\today}
\author{Barbara Piechocinska}
\affiliation{Uppsala University, Box 534, 751 21 Uppsala, Sweden}


\markboth{Wholeness as a Conceptual Foundation of Physical
Theories}{Barbara Piechocinska}

\pagestyle{myheadings}

\begin{abstract}

A description of physical reality in which wholeness is the
foundation is discussed along with the motivation for such an
attempt. As a possible mathematical framework within which a
physical theory based on wholeness may be expressed, elementary
embeddings along with the Wholeness Axiom are suggested. It is
shown how features of wholeness such as wholeness being
indescribable, more than the sum of parts, locally accessible and
giving rise to a self-similar, or holographic, type of order are
reflected in the mathematics. It is also shown how all the sets in
the mathematical universe may be expressed as emerging from the
dynamics of wholeness. Moreover, it is indicated how the
mathematics may be further developed so as to connect up with a
physical interpretation.

\end{abstract}

\maketitle

\section{Background and Motivation}

The purely reductionist approach to physics, in which the world is
modelled as being made up of separate identifiable parts that
interact, has proven successful over the centuries.  It was not
until the beginning of the twentieth century that experimental
data demanded a drastically new theory.  To accommodate the new
results the quantum theory was invented.  Despite the deeply
rooted reductionist tradition scientists discovered early on that
the quantum theory was not purely reductionistic, and yet it was
approached and formed from the classical side.  It is for instance
normal procedure to attempt to quantize classical systems, while
it could be argued that the quantum theory, being more fundamental
than classical theories, should be the starting point.

The new aspects, not reducible to a purely reductionist
perspective, were intimately related to what Bohr called
wholeness. From the invention of the quantum theory Bohr has been
clear to point out the key role that wholeness plays \cite{Bohr}.
When two quantum systems interact, or have interacted, we can in
general no longer view the system of interaction as being made up
of two separately existent parts. Instead an inseparable totality
is formed. The wave function of such an interacting system can, in
general, not be expressed as a product of the wave functions of
the two previously separated systems. And so a reductionist
picture based on the interactions of static identifiable parts
with definite properties cannot be maintained. The details of how
a nonlocal, indivisible, and dynamical form of wholeness plays a
central role are presented in the Ontological Interpretation of
Quantum Theory \cite{Bohm:UU} and a short summary of them will be
given in the Section 2. The effects of a correlation related to
the indivisible wholeness of quantum systems that transcends space
and time (entanglement) have been objectively demonstrated through
experiments \cite{Aspect}. Failing to conform to reductionism and
being essential in the foundations of the quantum theory, it would
be both intellectually satisfactory and of instrumental value in
future research to have an understanding of what philosophical
ideas wholeness entails.

Since the advent of the quantum theory severe problems have arisen
when attempting to form a consistent, unified description through
physics.  There appear to be incompatibilities between fundamental
physical theories such as the quantum theory and the theory of
general relativity. In attempting to find a common ground for both
of them various physicists have suggested wholeness
\cite{Bohm:WIO2} (chapters 5-6), \cite{Bohm:UU} (chapter 15),
\cite{Sachs}. The prospect of being a common factor in both the
quantum theory and the theory of general relativity hints at the
possibility of leading to a theory that transcends and includes
the both of them.

In view of the present incompatibilities, the difficulty to fit
contemporary physics into a reductionist framework, and the
indications that a non-reductionistic wholeness is essential in
physics, it is suggestive that we attempt a scientific description
that starts from wholeness. However, in order to engage in such an
attempt we are in considerable need of new ways of thinking
(fundamental concepts) as well as new mathematical tools, all due
to the particularly elusive nature of wholeness. This paper
addresses that need and presents a proposition for new foundations
of physics with the fundamental concepts and a mathematical theory
capable of representing them. In particular it proposes a concrete
mathematical description of Bohm's implicate order\footnote{see
Section 2.B for implicate order} that is based on a new axiom in
Set Theory.  It further proposes that the connection from the
mathematically described implicate order to physics be made
through the left-distributive algebra that this axiom generates,
and in particular through braids.

The paper is set up in the following way.  We shall start, in
Section 2, by summarizing the work of Bohm and Hiley in order to
give a specific example of how wholeness comes into the quantum
theory. Then we shall continue, in Section 3, by taking a closer
look at the concept of wholeness in general and at the particular
feature of wholeness that will serve as the fundamental concept
which will be represented mathematically. We will then proceed to
Section 4 in which an example of a mathematical framework capable
of expressing the fundamental concept and key features of
wholeness will be given. After that we shall, in Section 5, look
at some examples of how we expect the approach based on starting
from wholeness to be useful. In particular, we shall see how
wholeness facilitates the understanding of the properties of the
quantum theory, and indicate what mathematical structures (used in
physics) naturally arise from the proposed mathematics.  Finally,
in Section 6, we shall sum up the conclusions.

\section{Example of wholeness in the quantum theory}

When attempting to study physics in search of the underlying
concepts we find ourselves to be quite fortunate because there are
scientists who have already done a substantial amount of work in
this direction. Of particular interest is the work of Bohm, Hiley,
and collaborators.  A large part of their work originates in
intuitions deduced from Bohm's interpretation of the quantum
theory \cite{Bohm:52}, \cite{Bohm:UU}, \cite{Holland}. One of the
great advantages of this interpretation is that it allows for an
imaginative and intuitive understanding of quantum phenomena and
thereby opens up the door to further insights that may become
crucial for an extension of the quantum theory, possibly moving
into a new theoretical framework. In particular, through the
quantum potential, which as we shall see can be derived from the
Schr\"{o}dinger equation, this interpretation explicitly
accentuates the essential role of a dynamical, unbroken wholeness.
We shall now go into slightly greater detail in order to better
see the holistic aspect of physics and how they are suggestive of
Bohm's implicate order. The starting point will be Bohm's
interpretation of the quantum theory.

\subsection{Bohm's interpretation of the quantum theory}

From being a statistical theory about the outcomes of measurements
in which the actual phenomena involved cannot be analyzed
\cite{Bohr:analysis}(p. 72), Bohm developed an essentially
ontological interpretation of the quantum theory that provides
intuitive understanding and further insight into the actuality of
systems. In particular, by using the Hamilton-Jacobi theory, it
provides a framework within which classical and quantum effects
can be contrasted. Such comparisons offer a richer understanding
of what Niels Bohr called the unanalyzable wholeness \cite{Bohr}
and lead to the development of Bohm's implicate order
\footnote{Consequently, the implicate order can be seen as
providing a kind of understanding of the wholeness found in the
quantum theory that has the capability of going beyond the quantum
theory \cite{Bohm:implicate}.}.  It should also be mentioned that
aside from providing an ontology for, and an intuitive
understanding of the quantum theory, Bohm's interpretation also
gives an account for each individual quantum process and accounts
for measurements.

In Bohm's view, the wave function is not regarded as a complete
representation of a quantum state.  Its interpretation is not only
probabilistic, such as the square root of the probability of
finding a particle in a particular region. Instead the wave
function is an actual quantum field ($\psi$-field) that provides a
partial description of the formative cause, as well as a function
from which actual probabilities can be derived
\cite{Bohm:meaning}. This treatment of the wave function is what
eventually leads to an account of measurement.

In the original interpretation of Bohm \cite{Bohm:52}, a particle
is assumed to exist with a well defined position and momentum. One
can therefore say that the particle in fact moves along a definite
trajectory. The particle has a quantum field ($\psi$-field)
associated with it. They are physically inseparable.  On a deeper
level of the ontology the quantum field and the particle are
considered by Bohm to be different aspects of the same process
\cite{Bohm:UU}(sec.15.8).

Now Bohm resorts to Hamilton-Jacobi theory, which can be seen as a
theory of the interrelationships of rays and waves
\cite{Holland}(chapter 2). This treatment produces a picture in
which one can see how both classical and quantum effects arise. In
order to see this in the non-relativistic case, one starts with
the wave function written in its polar form
\begin{equation}
\label{eq:psi} \psi = R(\textbf{x}, t)e^{iS(\textbf{x}, t)/\hbar},
\end{equation}
where both $S$, the phase of $\psi$, and $R$, its amplitude, are
real fields that are dependent on the position, $\textbf{x}$, and
the time parameter $t$, and where $\hbar=h/(2\pi)$, $h$ being
Planck's constant. The next step is to derive the Quantum
Hamilton-Jacobi equation from the Schr\"{o}dinger equation.  Doing
this for one quantum particle one goes through the following
steps.  We start with the Schr\"{o}dinger equation, which
describes the time evolution of the quantum system,
\begin{equation}\label{eq:Schr}
i\hbar\frac{\partial \psi}{\partial t} =
-\frac{\hbar^2}{2m}\nabla^2\psi + U\psi,
\end{equation}
where $m$ is the mass of the particle and $U=U(\textbf{x}, t)$ is
a (real) classical potential. We now substitute (\ref{eq:psi})
into (\ref{eq:Schr}), work out the Laplacian, rearrange the terms
and divide them up into two equations, one for the real part and
the other for the imaginary part.  The real part then yields
\begin{equation}\label{eq:real}
\frac{\partial S}{\partial t} + \frac{(\nabla S)^2}{2m} + U + Q =
0,
\end{equation}
where $U$ is still a classical potential and $Q$ is the quantum
potential. More explicitly, we see that
\begin{equation}\label{eq:Q}
Q = -\frac{\hbar^2}{2m}\frac{\nabla^2R}{R}.
\end{equation}
When $Q\rightarrow 0$ we move from the quantum domain towards the
classical limit \cite{Holland}(p. 225) and for $Q=0$ equation
(\ref{eq:real}) becomes the classical Hamilton-Jacobi equation. If
we see the quantum potential as a type of potential energy we can
view (\ref{eq:real}) as an extended version of the conservation of
energy, valid in the quantum domain. In order to calculate the
momentum of the quantum particle we use the guidance condition
\footnote{It should be mentioned that an approach based on this
condition was introduced by other authors as ``Bohmian
Mechanics"\cite{Shelly:one} \cite{Shelly:two}. The mathematics of
it is in accordance with Bohm's own approach. However, several
parts, originally existing in Bohm's interpretation and crucial to
its philosophy (such as the quantum potential), have been omitted.
For a short comparison between the two approaches, see
\cite{Hiley:vaxjo}.}
\begin{equation}\label{eq:guide}
p = \nabla S
\end{equation}

The imaginary part of  (\ref{eq:psi}) in (\ref{eq:Schr}) is
\begin{equation}\label{eq:im}
\frac{\partial R^2}{\partial t}=-\nabla(R^2\frac{\nabla S}{m})
\end{equation}
This can be seen as an equation for the conservation of
probability.

The above approach grants us the unique opportunity of studying
the quantum potential, which is responsible for the non-classical
effects of the quantum theory.  If we can understand its nature
and physical relevance, we will come one step closer to
intuitively understanding quantum phenomena and possibly going
beyond them.

In order to pursue this we shall now stress how radically
different the quantum potential is from classical potentials.  Its
dynamics goes beyond what can be seen as a mechanical interaction
of external parts. To start with, we note that the quantum
potential has no external source from which it can be seen to
emanate, such as for instance the gravitational potential could be
seen to emanate from a massive body in space. This is basically
due to the fact that the quantum potential is constructed from the
$\psi$-field, which itself depends on the whole system and lacks
an external source. And so, the quantum potential's dependence on
the $\psi$-field introduces an irreducible dependence on the
entire environment. When more particles are involved it becomes
even more clear that the wave function depends on the whole
system. Here we encounter a holistic aspect in which wholeness is
more than just all the parts and their interactions.

Moreover we note from equation (\ref{eq:Q}) that the quantum
potential has a different form than classical potentials.  We find
$R$ both in the numerator and denominator, which means that the
quantum potential does not in general necessarily diminish with
distance. Even though the $\psi$-field may go to zero as the
distance increases, the quantum potential does not have to
diminish. This means that remote features may have significant
influence on particle movement. When more particles are involved,
this can give rise to the phenomenon of non-locality.  The quantum
potential will in general depend on the positions of all $N$
particles,
\begin{equation}
Q = -\frac{\hbar^2}{2m}\frac{\nabla^2R(x_{1}, x_{2},...,
x_{N})}{R(x_{1}, x_{2},..., x_{N})},
\end{equation}
so that the quantum force $-\bigtriangledown_{k} Q$ on some
particle $k$ depends on all particles.

Furthermore we note that it is not the intensity of the
$\psi$-field that regulates the effect it has on the movement of a
particle but its form.  This again is due to the presence of $R$
in the denominator.  An increase in the amplitude, by some factor
$c$, has no bearing on the quantum potential, as $c$ divides out.
This in turn suggests that the action of $\psi$ on the particle,
through $Q$, is of a different kind than classical pushing or
pulling (a mechanical transfer of momentum and energy), and is
thoroughly discussed by Bohm and Hiley \cite{Bohm:UU}. They
suggest that the quantum field be treated as an information
potential. The usage of the word information is different
\cite{Hiley:vaxjo} from Shannon's information \cite{Shannon},
which refers to our ignorance or certainty about a system.
Instead, Bohm and Hiley's information ``in-forms," actively puts
form into. The information in the quantum potential guides a
particle along its way.  It is information relevant to the
objective movement of a particle, information that has meaning to
the particle. When it guides a particle it is referred to as
``active information." Bohm and Hiley have suggested the metaphor
of a boat being guided by the radar.  The actual waves of the
radar do not push or pull the boat, as do the waves of the sea. In
a similar way the quantum potential guides through information
that is meaningful to the particle, by putting form into the
motion and not by pushing it.

\subsection{Dynamical unfoldment and enfoldment and the implicate order}

The dynamics suggested by the quantum potential clearly indicates
a radical change from the classical framework. The quantum
potential appears to suggest a dynamics where the irreducible
totality of a system and its environment acts (from beyond any
particular spatial source) and forms the explicit movement of some
particle. In other words, the lack of an external local source for
the quantum potential along with its irreducible dependence on the
whole system and, in general the existence of non-locality,
suggests that it operates from an irreducible order beyond space
and time. Bohm called this the implicate order.  The implicate
order is dominated by its holistic aspects and can be seen as a
dynamical totality where things cannot be distinguished from each
other and are instead inseparably intertwined. In other words,
despite a lack of crystallization, or definite boarders and
distinguishability (that we are used to seeing in space), all the
relationships are intact and preserved.

Operating from the implicate order the quantum potential gives
explicit form to the movement of a particle.  Here we can talk
about the explicate order.  The explicate order is the order where
things can be distinguished from each other, or seen to exist
outside of each other, and approximated as independent.

An important fact here is that in order to form the movement of
the particle the $\psi$-field needs information from the whole
environment that is processed as a totality.  We see how the
implicate order, which is beyond the visible (explicate), plays an
essential, formative role in the dynamics. And so, the movement of
the part comes out of the totality, not the other way around.
Since the totality is irreducible we cannot reduce it to parts.
This is how wholeness is greater than the sum of its parts and
needs to be considered in any serious attempt to go beyond the
quantum theory.  This is the fundamental idea upon which we will
continue to build and why we shall try to derive the movement of
parts from the dynamics of wholeness instead of starting with
parts and letting their sum describe a reductionistic wholeness.

To do this we shall need to discuss the nature of the process of
going from the implicate order to the explicate order and the
other way around.  This will be done in terms of unfoldment and
enfoldment. The dynamical concepts of unfoldment and enfoldment
were introduced and developed by Bohm \cite{Bohm:WIO2} in an
effort to provide a common basis capable of accounting for both
quantum and relativistic effects. The implicate order is the
enfolded order while the explicate order is the unfolded order. To
explain these aspects Bohm proposes an illustrative metaphor. One
places a cylinder container with a smaller radius inside of a
fixed one with a larger radius.  Then one pours glycerine between
them.  On top of the glycerine one then places a drop of ink.  If
one at this point starts turning the inner cylinder the ink-drop
will smear out and become a line. After a while no ink will be
visible.  This is analogous to the enfoldment process.  The
ink-drop was first explicit and has now become enfolded into the
order of the molecules and is implicit. Though the drop cannot be
seen explicitly it is still there, implicitly in the order.  If we
now start turning the inner cylinder in the other direction the
ink-drop will appear again. This is then analogous to the
unfoldment process.  The ink-drop becomes explicit again.  This
example descriptively expounds the ideas of unfoldment and
enfoldment but should not be taken literally.

The idea that Bohm tried to convey was that every thing is in some
sense enfolded into the whole and that the whole is unfolded into
every thing.  Bohm called this dynamics the holomovement which is
a holistic pulsation in which orders unfold and enfold. Emphasis
should be put on the fact that this fundamental process is not a
movement within space-time (like in the example of the ink-drop)
but rather a process in which ultimately space-time and its
contents are created\footnote{To see how the contents are created
one extends Bohm's interpretation to the quantum field theory
where a super-quantum potential affects the field equations in a
non-local and non-linear way.  There, it is not only the motion of
a particle that is formed but the very existence of the particle
itself.}. Recall that the quantum potential does not operate from
any particular part in space but originates in an irreducible
totality, and yet is locally accessible from any point within
space where it puts explicit form into the movement. Likewise, the
fundamental process originates in the implicate order which
unfolds into an explicate order and forms it through the
unfoldment.  Because of the wording one might be inclined to think
that there exists an implicate order and an explicate order and
that they interact through unfoldment and enfoldment.  This,
however, is not what is being proposed here. Instead, focus should
be put on the process, or movement.  It is the nature of the
unfoldment and enfoldment that is such that one can see it as
giving rise to the explicate from the implicate. The explicate
order does not have a permanent and independent existence. It is
continually being created and dissolved. And the existence of the
implicate order is of a subtle kind, it cannot be said to exist
explicitly because it is implicit and beyond space and time.
Therefore, it is suggested, movement is fundamental and can be
expressed in terms of the enfoldment and unfoldment.

It has further been argued by Bohm and Hiley, in \cite{Bohm:UU}
(chapter 15.3), that essentially all the quantum mechanical laws
of movement can already be seen as unfoldment and enfoldment.  The
value of a wave function at a particular time and point in space
depends on the whole space at previous moments. Consequently a
particular region can be said to enfold contributions from the
whole space. Then, as it evolves, it unfolds into the whole space.
The authors also show how Huygen's principle and Feynman diagrams
can be understood in terms of enfoldment and unfoldment.  Huygen's
principle tells us that the propagation of a wavefront can be seen
as each point on the wavefront acting as a wave-emitting source
point.  In other words, each point enfolds contributions from all
other points on the wavefront and then unfolds, giving rise to a
new wavefront.  Since Feynman diagrams can be considered as based
on Huygen's principle \cite{Bohm:UU}(p. 355), enfoldment and
unfoldment can be considered as fundamental for them as well.

Summarizing Section 2 we may say that the essential concept that
can be extracted from Bohm and Hiley's work is that there is a
dynamical, unbroken wholeness. The dynamics seems to consist of
unfoldments and enfoldments of explicate and implicate orders.

\section{Description from wholeness}

\subsection{The concept of wholeness}

Starting with wholeness presents a particular challenge for a
physical description of reality.  This is due to its indefinable
and elusive nature.  Therefore, the approach taken here will be to
assume the existence of an indefinable wholeness and investigate
what conclusions this assumption may lead to in terms of physical
theories.  However, before making such an assumption some features
of wholeness will be highlighted.

Wholeness is taken to be the ontological basis for the physically
measurable reality. It is viewed as the totality that implicitly
unites and gives rise to everything observable, but is more than
the observable. In particular we will be dealing with a wholeness
that is larger than the sum of its parts. Working with parts
towards wholeness not only leaves us with an insufficient
\cite{Primas} and severely limited picture of wholeness, but also
with one that is inconsistent\footnote{Publications regarding the
whole being more and different from the parts can be found in
\cite{Anderson}, \cite{Primas}.}. Furthermore, we do not wish to
confine ourselves to working with parts because we are interested
in describing a wholeness that is similar to, or at least not more
limited than, the wholeness found in the quantum theory (see
Sections 1 and 2).

Another central aspect of wholeness is that it is indefinable.
This can be intuitively understood in the following way.  Being
whole is not in opposition to anything, not even to part. If
wholeness were opposed to part and thereby excluded part it would
not be whole. And so, it is not possible to gain a true
understanding of wholeness based on division, or differentiation,
and opposition. Consequently a description of wholeness is not
possible in terms of our language because it is based on
opposition and differentiation \cite{Hegel}, \cite{Bohm:WIO2}
(chapter 3), \cite{Bohm:Peat}. Differentiation and opposition lead
to definitions and to information that is based on the concept of
lack of knowledge. To define something is to delimit it and say
that it is that which is found within the limit and not that which
is found outside of the limit.  Such a delimitation cannot be
performed on wholeness and one can therefore say that wholeness is
not definable. In fact, trying to describe wholeness may be
likened to attempting to reach infinity by, say, counting all the
natural numbers, 1,2,3,... No matter what number we reach it will
not be infinity.  Just like infinity is not a number, wholeness is
not a part, or reducible to a description. As a consequence we see
that any description is a limitation. This must also hold for a
strictly holistic view of wholeness where it is seen as undivided
and forever non-separable into individual modes. In what follows
we shall therefore take the broader view that the holistic
perspective and the reductionistic perspective are complementary
in the sense that both are required in a description of reality
based on the assumption that there is wholeness.

The relationship between the holistic and reductionistic aspects
can be described as according to a self-similar, or holographic
type of order (the order of Bohm's implicate order as described in
the previous section). To see the self-similarity and holography
we must first recognize that the ontological basis of each part is
the holistic wholeness. In other words, one part implicitly
contains all parts at its ground and all parts implicitly contain
that one part, at their ground. Wholeness is in each explicit
part, as that parts innermost nature or ontological ground, while
from the implicate (holistic) perspective each part is inseparable
from wholeness. And so, wholeness is (in) each part and each part
is (in) wholeness. This can be seen as self-similar because
through wholeness there is a similarity between a part and all
parts, and holographic, because each part implicitly contains
information from all parts.  This kind of order is a consequence
of the holistic properties and is already found in the quantum
theory, where we see how the dynamics of a quantum system has an
irreducible (holistic) dependence on its surroundings
\cite{Bohm:UU}.

One should bear in mind that wholeness is not some physically
measurable quantity and its existence can therefore not be
directly, physically, verified through objective experiments.
Instead it is the source of the physically measurable quantities.
An analogy can be made to superposition in the quantum theory,
which can never be explicitly measured. As described above,
wholeness is qualitatively different from part or thing for
instance in that it cannot be explicitly defined. However, we
shall see that if wholeness is assumed a consistent picture of
reality will take form in which wholeness, through its nature, may
be seen as the source of physical reality. Perhaps just like the
acceptance of infinities in mathematics provides a rigorous
foundation for and definition of real numbers, an assumption of
wholeness might account for the existence of the physical
universe.

In order to provide a rigorous mathematical description for a
physical theory that assumes the existence of wholeness we will
need to find and specify what we consider to be fundamental
concepts.  Then, we will need to find a way of mathematically
expressing them. Since the existence of wholeness is assumed, it
is vital that this be reflected in the mathematical description on
a fundamental level. We will now quickly review our understanding
of a dynamical wholeness in order to provide the fundamental
concept.

\subsection{The fundamental concept}

The fundamental concept, upon which we shall base our description
of the physical universe is movement, or as suggested by Bohm
\cite{Bohm:WIO2}, process. Besides the motivation provided by
Bohm's approach (in Section 2) this can be motivated in the
following way. In our description of the physical universe we wish
to take wholeness as fundamental. This requires our description to
be complementary. We need both the holistic perspective and the
reductionistic perspective. Although a description in terms of
parts is certainly possible in the reductionist spirit, it cannot
be a fundamental base for a description from wholeness that
includes the holistic perspective.

Here is a slightly different way of seeing why parts may not be
the best way of describing physical reality from wholeness.  For a
long time physicists have been trying to find a fundamental
building block, entity, or part. Let us assume that such a thing
exists and see what properties it should possess. However, let us
not limit ourselves to building blocks that have positions or
extensions in space or time. Suppose we find two such fundamental
building blocks, then we will find that, the blocks being truly
fundamental, we cannot make any kind of distinction between them.
If no such distinction exists (not even a distinction in time or
space) we conclude that the two fundamental building blocks are
one. Consequently, we can say that there is only one fundamental
building block for everything.  This can be interpreted in two
ways.  The fundamental building block is either nothing or it is
wholeness.  Neither of these cases is consistent with a
reductionist framework where identifiable parts are taken as
fundamental.  One reason is that neither nothing nor wholeness
have outer limits which would allow us to define or identify them.
Instead they are thing-less.  And so, they may not be considered
as parts.

As parts do not meet our expectations in terms of fundamental
concepts for a description from wholeness we turn to movement or
process as a plausible alternative.  This movement or process
should in its totality not be viewed as confined to physical space
and time and is not reducible to a description of some parts that
move in a sequential manner. Instead it is the fundamental
dynamics of wholeness, and that which through self-limitation
gives rise to physical space and time along with parts which may
be viewed as properties of, or invariances in, the movement.  And
so, the dynamics is such that any part, and the change of any
part, may be traced back to the fundamental movement. Such a
dynamics needs to be wholeness preserving. Here follows an
explanation. Wholeness may be said to be a totality implicitly
containing everything. This in turn appears to imply that for it
to be truly whole it cannot change. If it were to change and
become different, then this would imply that it was not truly
whole in the first place because it did not contain itself after
the change. At the same time we know that change, at least in our
physical world, exists. Therefore, we are looking for a dynamical
description that allows wholeness to remain essentially unchanged
though moving, from the perspective of parts, and thereby allowing
for change. We plan to capture this dynamical feature of
wholeness, the wholeness preserving movement, mathematically in a
fundamental way.  This will be the fundamental concept. And so
essentially, the proposition here is to describe our physical
reality as an expression of the dynamics of wholeness.

Furthermore, we should like to have a way of expressing
diversification, preferably as a result of the wholeness
preserving movement.  We also expect to have the possibility of
expressing time, space, physical objects as properties of or
invariances in the movement.

\section{Mathematical framework}

In order to provide a mathematical foundation that is capable of
expressing features of wholeness at the very fundamental level, it
is suggested that we start with set theory, since most of
mathematics can be expressed in terms of it, and add the Wholeness
Axiom introduced by Corazza \cite{Corazza} to the normal axioms of
set theory. The normal axioms of set theory are the
Zermelo-Fraenkel set of axioms and the Axiom of Choice,
abbreviated ZFC \cite{Zermelo:1908}, \cite{Fraenkel}, \cite{ZFC}.
For approximatively eighty years it has been known that most of
mathematics is derivable from them. These axioms refer to sets and
are formulated in the formal language $\{\in\}$. This means that
the only relation used in the formulation of the axioms is the
membership relation, $\in$.

The mathematical theory within which we shall attempt to describe
a theory from wholeness will contain the ZFC set of axioms
together with the Wholeness Axiom and will be referred to as
ZFC+WA. We will now proceed to see what the Wholeness Axiom is and
what features of wholeness ZFC+WA is capable of expressing.

\subsection{The mathematical universe $V$}

An important concept for our purposes is the mathematical
universe. The mathematical universe, $V$, is a proper class and
can be expressed as:
\begin{equation}\label{V}
  V=V_{0}\bigcup V_{1}\bigcup V_{2}\bigcup...,
\end{equation}
where the different stages are $V_{0} = \emptyset$, $V_{1} =
P(V_{0})$,$V_{2} = P(V_{1})...$, with $P()$ denoting the power set
(the set of all subsets).  It is interesting to note that there is
a salient analogy between the mathematical universe $V$ and
wholeness.  As previously mentioned, wholeness implicitly contains
everything but is more than the sum of the parts, and is itself
not an explicitly existent entity (a part). The nature of its
existence is implicit. In a similar way $V$ may be said to contain
all the mathematically existent parts and yet be more than their
sum and not itself an explicitly, mathematically existent entity.
Saying that V is the union of $V_{0}, V_{1}, V_{2},...$ is like
saying that wholeness is, or contains, all the parts. The
existence of each stage, $V_{x}$ for any ordinal $x$ in $V$,
follows from the ZFC set of axioms. It is interesting to note,
however, that the existence of $V$ itself cannot be shown from the
ZFC set of axioms, since $V$ is not a set. In other words, $V$ in
its totality cannot be shown to be an explicitly existing
mathematical object.   The fact that all its stages can be shown
to exist seems to imply that $V$ should exist as well, however,
its existence cannot be explicitly shown.

It should be stressed that our description of $V$ is but a
description of a feature of wholeness and far from a complete
description of wholeness. Such a description cannot be expressed
in explicit terms because the nature of wholeness is not reducible
to the explicit.  Our description in terms of $V$ only refers to
the explicit parts and states that it contains all parts but is
more than that.  This way of referring to wholeness is employed
because the aim of this theory is to account for the existence and
dynamics of all explicitly existent parts of reality.

\subsection{The wholeness preserving movement, $j$}

As argued earlier, one of the fundamental concepts upon which we
wish to base a theory from wholeness is the wholeness preserving
movement. This means that we wish to postulate the existence of
such a movement in order to enable a prediction of quantifiable
characters of an experimentally verifiable entity in terms of the
wholeness preserving movement. It turns out that the Wholeness
Axiom can be seen as expressing the wholeness preserving movement.
The Wholeness Axiom \cite{Corazza} is the assumption that there is
a nontrivial elementary embedding, $j$, from the mathematical
universe, $V$, to itself:
\begin{equation}\label{WA}
  j: V \rightarrow V
\end{equation}
For $j$ to be an elementary embedding means that if some formula,
$\phi(X)$ holds true in the domain, which for $j$ is $V$, then
$\phi(j(X))$ must hold true in the codomain, which is also $V$ in
this case.  In other words, $j$ is reflective and truth
preserving.  It can be said to be wholeness preserving because due
to its non-triviality it actually changes parts while preserving
all structures in $V$.

Definition (\ref{WA}) can be axiomatized in a consistent
manner\footnote{despite Kunen's theorem\cite{Kunen}} so that
ZFC+WA is obtained. This is done by adding $j$ to the language
$\{\in\}$, so that we have $\{\in,j\}$ and saying that the ZFC set
of axioms is valid with the addition that all instances of
Separation and no instance of Replacement\footnote{The Separation
and Replacement schema are axioms contained in the ZFC set of
axioms.} are valid for $j$, and the addition of axioms that
express $j$ as a nontrivial elementary embedding from $V$ to
itself (Nontriviality axiom and Elementarity axioms). The meaning
of this in terms of wholeness will be discussed in the succeeding
sections.

The Wholeness Axiom was developed to prove the existence of all
large cardinals, in mathematics.  And so, elementary embeddings,
such as $j$, are closely connected with infinities.  One of the
aspects they have in common with infinities is that they disclose
precisely the self-similar, or holographic, type of order we
attribute to wholeness and find in the quantum theory.  An example
of self-similarity for the real numbers is that there are as many
real numbers in, say, the interval between $0$ and $1$, as there
are on the whole line. Elementary embeddings can be seen as
displaying perhaps the ultimate form of self-similarity because
one cannot distinguish a part from the whole with any definable
formula.

\subsection{The critical point $\kappa$}

The introduction of the wholeness preserving movement,
mathematically formulated as the Wholeness Axiom, brings about
some interesting consequences. Of primary importance is the
arising of a critical point. This happens because the axioms for
$j$ ensure that there is a least ordinal moved by j.  This ordinal
is denoted $\kappa$ and is called the critical point of $j$.  In
particular, we see that Nontriviality for $j$ asserts that there
has to be some set $X$ for which $j(X)\neq X$, and so, there has
to be some smallest ordinal for which this condition holds true.
The restriction of $j$ to any set of a rank less than $\kappa$ is
the identity. In other words, for any set $Y$, of rank less than
$\kappa$, $j(Y)=Y$. Being an elementary embedding, $j$ is truth
preserving. This means that properties, operations, and relations
that hold true for some sets $X_{1}, X_{2},...\in V$ also hold
true for $j(X_{1}), j(X_{2}),...\in V$. This property of the
elementary embedding makes it possible for us to say more about
$\kappa$.  It turns out, for instance, that $j(\kappa)>\kappa$
(see the Appendix) and that $\kappa$ is an infinite cardinal with
all the large cardinal properties \cite{Corazza}.  In can be
mentioned that, as described in Section 4.E, thanks to $\kappa$,
$j$ contains a lot of creative or generative power. This might
become useful for physical theories.

\subsection{Separation and no Replacement for $j$}

Recalling and recapitulating some features of wholeness we can say
that wholeness is indescribable, it is more than the sum of parts,
and yet it is locally accessible and that through which every part
emerges. All of these features are represented in ZFC+WA.  In
order to see this, it is of interest to note that ZFC+WA does not
restrict wholeness, or the wholeness preserving movement, to some
explicitly existent mathematical object. All it does along those
lines is to assume the existence of a wholeness preserving
movement. The wholeness preserving movement is itself never
defined by some specific formula.  In fact, it cannot be defined
within set theory. $j$ is neither a set nor a proper class.  The
fact that $j$ is not a set follows from $j$ being defined on all
of $V$. To see why it is not a proper class either we need to take
a look at the axioms of Separation and Replacement.

When axiomatizing the wholeness preserving movement, $j$, it is
said that all instances of Separation and no instance of
Replacement are valid for $j$. Given a set, $A\in V$, Separation
allows us to talk of a subset of $A$ in which some property is
true for all the elements of that subset.  So, $j$ having all
instances of Separation means that for all properties, $P$,
depending on $j$, we can take any set, $A$, and look at the subset
of $A$ containing all elements for which $P$ is true. That subset
of $A$ is itself a set. Separation for $j$ makes $j$ interesting,
powerful, and promising for a further development in terms of a
physical theory, because it allows $j$ to act locally, meaning on
any particular set. This allows us to use $j$ in local
descriptions. What $j$ does on the entire mathematical universe,
it also does locally on any set.

Replacement, on the other hand tells us that for any set $A$ and
any rule that associates with each element $x$ of $A$ a set
$Y_{x}$ there is a set $B$ that consists precisely of all $Y_{x}$,
where $x\in A$. Replacement is useful for class functions.  Class
functions are functions that are not sets but proper classes
defined on some proper class, such as for instance on $V$.
Replacement guarantees that the range of a class function being
restricted to some set is also a set. The range being a set makes
its rank limited.  So, we can say that Replacement makes sure that
there can be no definable way of going through the top of $V$ in a
certain amount of steps (indicated by the rank of the domain). No
instances of Replacement for $j$ tells us that for all functions,
$F$, depending on $j$, there may be no set consisting of all the
$Y_{x}'s$ associated with the elements $x$ of a given set. Indeed,
Separation for $j$ can be used\footnote{The proof makes essential
use of Kunen's theorem, see Proposition 3.6 in \cite{Corazza}, and
\cite{Kunen}} to show that if $F$ is defined by letting
$F(0)=\kappa$, $F(1)=j(\kappa)$, $F(2)=j(j(\kappa))$, etc., and
letting $F(x)=\emptyset$ for any $x$ that is not a natural number,
then the restriction of $F$ to the set of natural numbers has a
range that goes all the way through $V$.  More simply, the
sequence $\kappa<j(\kappa)<j(j(\kappa))<...$ extends above every
rank in the universe.  One says that the 'critical sequence for
$j$ is cofinal in the mathematical universe'. Hence $F's$
depending on $j$ do not exist as sets or proper classes in $V$ and
may be capable of going through the top of $V$. No Replacement for
$j$ assures us that $j$ is not a proper class because it is not
definable by a formula within set theory. If it were definable in
set theory, then by ordinary Replacement in set theory, the $F$
defined above would have the property that the range of $F$
restricted to the natural numbers would be a set. As observed
above, this is not the case.

 Going back to our previously mentioned features of wholeness,
$j$ is indescribable and more than ``the sum of the parts". At the
same time it acts locally and is present in every part of V.

\subsection{Laver sequences--emergence of all sets in V}

Taking the wholeness preserving movement to be fundamental it is
desirable to have a way of seeing how the movement gives rise to
every part. Is there a way of describing the emergence of all sets
in the mathematical universe through the wholeness preserving
movement? Although, strictly speaking, mathematics is in general
not viewed as emerging from anything, it is possible to see all
the sets in the mathematical universe as an expression of the
dynamics of wholeness. In order to see how this can be done we can
use Laver sequences \cite{Laver}. The definition of a Laver
sequence given here is generalized by, and due to, Corazza. A
Laver sequence is a set $S$ of length $\kappa$, $\kappa$ being the
critical point of $j$, that has the property that for any set $X$
in $V$ there is an elementary embedding $i: V_\alpha\rightarrow
V_\beta$ such that the $\kappa$:th term of $i(S)$ is $X$. In other
words, every set in the mathematical universe can be located as
the $\kappa$:th term of an image of the Laver sequence $S$ by some
elementary embedding $i$.  Moreover, all such $i$'s are derived
directly from $j$ itself.  Therefore, every set exists as if in
seed form in the single point $S$; its existence as a set becomes
apparent when the appropriately derived embedding $i$ is applied
to $S$. So, given $\kappa$ we can construct a sequence such that
every set in the mathematical universe can be seen to emerge from
it through movement.  The existence of such sequences can be shown
assuming the Wholeness Axiom.  Another way of looking at Laver
sequences is to see them as functions $f: \kappa\rightarrow
V_{\kappa}$ such that for any set $X\in V$ there is an elementary
embedding, $i: V_\alpha\rightarrow V_\beta$, such that
$i(f)(\kappa)=X$.  As functions, Laver sequences live in the
mathematical universe and may be said to be made out of parts.
However, a detailed understanding of the parts does not facilitate
the understanding of them.  It is not until one ``shines" an
elementary embedding on them that they become ``active" and one
realizes their power to give rise to all sets. These global
properties, not perceptible from the details but visible through
movement, are what make them interesting.  These properties may
come to play a key role in finding new formative causal structures
\cite{BH:fusion} for physical theories from wholeness.

In short, the Wholeness Axiom provides wholeness preserving
movement. As a consequence there is a $\kappa$ with which we can
form a Laver sequence from which all sets in $V$ can be seen to
emerge through the movement.

\section{From the Wholeness Axiom towards physics}

We shall now sketch out a possible way of elaborating on and
developing the approach of starting from wholeness and its here
introduced mathematical formalism.  However, before going into the
further development of the mathematics, we note that this approach
already provides us with a conceptual structure that makes it
possible to evolve our intuitive understanding of the theories we
already have, such as the quantum theory.  Let us take a look at
such an example.

\subsection{Insights into Heisenberg's principle of
indeterminacy}

Here we argue that if an undivided wholeness is assumed, then the
holistic view can provide insights into features of the quantum
theory along with intuitive understanding. In order to better
understand what a holistic view entails and adds, we shall, at
this point, examine some fundamental differences in the concepts
and requirements for reductionism and holism.  A good description
of these differences may be found in \cite{Bohm:QT}.

Reductionism presupposes the possibility of decomposition into
identifiable parts.  This means that we view the system that we
study as essentially being made up of separate, identifiable
parts, that interact with each other as described by laws.  The
description of the whole system is reducible to a description of
its parts.  In order to be able to have identifiable parts we need
to have both strict causality and strict locality (or space-time
continuity).  And having both means that we can in principle
always find identifiable parts.  In other words, identifiable
parts are equivalent to strict locality together with strict
causality.  Should either of these fail, then we will no longer be
able to identify parts.

Let us now examine the holistic view.  This involves using
indivisible wholeness as most basic and fundamental.  The holistic
view assumes that there are no separate parts.  Although there may
be distinguishable modes on some level, on a deeper level we will
see that separation and precise distinction are not possible.  In
other words, on that level we would not be able to find
identifiable parts.  As a consequence we could not have both
strict causality and strict locality.  We could have one of them,
but that would imply that we would not be able to have the other.
As we saw in the second paragraph, having both implies
identifiable parts. Another alternative is that we could have
neither strict causality nor strict locality. It is interesting to
note that this is exactly what happens in the quantum theory
\cite{Bell}, and can there be seen as an expression of
Heisenberg's principle of indeterminacy.  One can intuitively
think of position and time as expressing the local aspects, while
energy and momentum can be considered to express the causal
aspects \cite{Bohm:QT}. Assuming wholeness, the appearance of a
principle such as Heisenberg's principle of indeterminacy can
hereby be seen as a necessary consequence of the undivided
wholeness itself. We can see why strict causality and strict
locality are not simultaneously possible.

\subsection{Connection of the mathematical framework with physics}

Let us now continue by outlining, step by step, a possible way of
further developing the mathematics so as to establish a
correspondence with physical phenomena. A detailed exposition of
the following presentation will be published elsewhere by the
author.

Since physics is about describing how different processes relate
to each other we are interested in seeing the kinds of processes
that the wholeness preserving movement can generate and what
relational structure this gives rise to. A way of generating
different processes is to have $j$ interact with itself.  To
describe this mathematically we define the binary operation of
application, $\cdot$, \cite{Corazza:spectrum}(page 54), and apply
$j$ to itself. In doing so repeatedly, we end up with a universal
algebra $A_j$ comprised of elements such that each element is
obtained by finitely many applications of $j$. Thus $j$,  $j\cdot
j$,  $j\cdot (j\cdot j)$,  $(j\cdot j)\cdot j$ are some of the
elements found in $A_j$.  $A_{j}$ together with application is a
free left-distributive algebra with a linear ordering on it
\cite{Laver:freeness}. Left-distributivity means that if $a$, $b$,
and $c$ are elements of $A_{j}$ then $a\cdot(b\cdot c)=(a\cdot
b)\cdot(a\cdot c)$. These kinds of algebras on elementary
embeddings have originally been studied by Laver
\cite{Laver:freeness} and an introduction to them can be found in
\cite{Dehornoy}.

With $A_j$ we now have a process algebra where the processes can
act on each other through application and every process is
obtained from the wholeness preserving movement. We note that
conceptually, this algebra fits exceptionally well into the
process algebra approach advocated by Bohm and Hiley
\cite{Bohm:UU}, \cite{Bohm:WIO2}. Their goal is to describe nature
in terms of process that originates in the dynamical holomovement
of the implicate order \cite{Bohm:WIO2}. The algebra $A_j$ on the
elementary embeddings is precisely that. It is a process algebra,
where all the elements are elements of movement generated by the
wholeness preserving movement.

The next step is to note that the process algebra $A_j$ is,
through its left-distributive structure, directly tied to braids.
Left-distributivity can be said to be behind the geometry of
braids \cite{Dehornoy}. It turns out that the process algebra
$A_j$ is isomorphic to special braids \cite{Dehornoy}(p 103).
Special braids are obtained by letting the unit braid, 1, act on
itself through braid exponentiation \cite{Dehornoy}(p 28). Every
braid in the braid group can be decomposed into shifted products
of special braids. By looking at $A_j$ as the algebra of special
braids we focus our attention on the structural aspect, that is,
how processes are related to processes (as is usually done in
physics), while steering away from the insight of what these
processes really are, elementary embeddings representing the
wholeness preserving movement whose innermost nature is wholeness.

In physics braids do not only turn up in certain applications like
the experimentally verified statistics of anyons in the Quantum
Hall effect \cite{Prague} or more theoretically as Yang-Baxter
operators providing useful solutions for low dimensional quantum
field theory \cite{Yang:field} or lattice statistical physics
\cite{Baxter}, but also in the very fundamental structure of
physics itself as in, for instance, the form of braided
commutativity \cite{Majid}. In fact, some mathematical physicists
are studying braids and Hopf algebras are doing so in the hope of
unifying the quantum theory with general relativity \cite{Majid}.

Just to give a bit more insight into how braids can be related to
physics in the aforementioned approaches we follow with a few
comments. One particular connection between braids and statistical
mechanics \cite{Wu} is usually described in terms of knots, or
more generally links.  However, a link can always be written in
terms of braids, as the closure of a braid. Links that may be
transformed into each other without the tearing of any strings,
that is using the so called Reidemeister moves
\cite{Reidemeister}, are said to be equivalent. It turns out that
a link can be seen as a lattice on which we might, for instance,
place spins either at the lattice sites (spin models) or at the
edges (vertex models).  The procedure is then to construct models
such that all lattices corresponding to equivalent links give us
the same partition function.  This way the partition function is a
link invariant.

However, as already mentioned the connection is not limited to
statistical mechanics.  Depending on features such as global
properties and discreteness, the topology of braids can also be
connected with quantum features. In fact, the connection between
statistical mechanics and knot theory was first discovered by
Jones \cite{Jones} by noting the similarities between von Neumann
algebras and the braid relations. A von Neumann algebra is a
$C^{*}$-algebra, used in algebraic approaches to the quantum
theory such as \cite{Haag}. Jones derived a link invariant, called
the Jones polynomial, which through the connection with von
Neumann algebras can be seen as providing expectation values.

The connection between knots and physics has been further explored
by Kauffman \cite{Kauffman}. And here we wish to especially point
to a certain bracket polynomial \cite{Kauffman}(p 97), which can
be defined for links or braids. The approach taken by Kauffman in
general is to start with a braid (or link) and with the help of
his bracket arrive at an algebraic structure, a link invariant,
which can take on physical interpretations such as expectation
values. What is proposed here is an approach more along the lines
of Dehorony \cite{Dehornoy} where we not only look at the
algebraic structure associated with a particular braid, but the
structure associated with all possible braids, and in fact from
where they can be seen to originate (a structure like the process
algebra $A_j$). Applying structures such as the bracket to $A_j$
would then provide a physical interpretation. Hence the idea is
that in so doing, this approach would be able to provide a
background for the many already discovered connections between
braids (links and knots) and physics, as well as possibly lead to
new discoveries.

The outlined proposition is schematically depicted in figure 1.
\begin{figure}
  \centering
  \includegraphics[width=8cm]{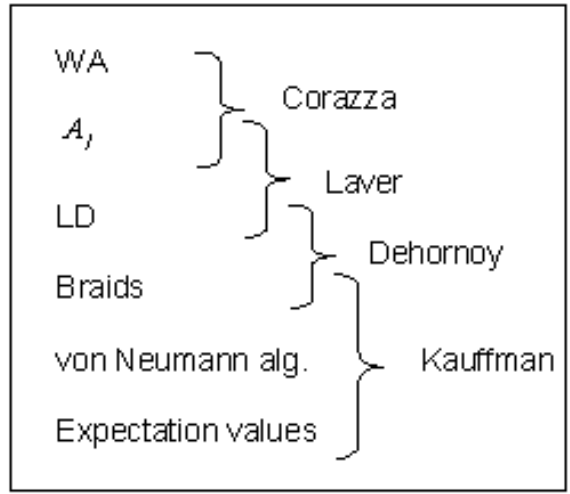}
  \caption{Schematic picture summarizing a possible development of
   the approach from Wholeness.  The Wholeness Axiom (WA) gives us $j$, which generates
   the universal algebra $A_j$, which is left-distributive (LD) and behind the geometry
   of braids, which through Kauffman's bracket can be connected to statistical mechanics and the
   quantum theory.  The names on the right refer to mathematicians who in their work (see References)
   in detail describe the alluded connections.}\label{fig1}
\end{figure}

\section{Conclusions}

In conclusion, it is suggested that a dynamical wholeness may be
capable of providing a conceptual foundation for the development
of physical theories. By starting out from a broader perspective,
primarily motivated by the quantum theory, it provides the
guidelines for the creation of a theory that may be valid where
other currently existing theories reach their limits. As a
fundamental concept upon which to build a theory from wholeness we
suggest a wholeness preserving movement. Such a movement may be
mathematically expressed in a fundamental way if the Wholeness
Axiom is added to the currently established axioms of Set Theory.
Such a mathematical foundation naturally allows for a description
of the emergence of all parts of the mathematical universe. It
also contains holistic features as indescribability, more than the
sum of parts, local accessibility and the holographic type of
order that we find in the quantum theory.

We have seen an example of how starting from wholeness facilitates
the understanding of seemingly paradoxical features of existing
theories.  We have also indicated how the approach may be further
developed mathematically, through a left-distributive algebra and
braids, so as to connect up with physics.

\section*{APPENDIX: Proof that $j(\kappa)>\kappa$}

Let $j: V\rightarrow V$ be a non-trivial elementary embedding from
the mathematical universe to itself, with $\kappa$ as its critical
point.  Since $j$ is by definition non-trivial $j(\kappa)\neq
\kappa$.  Let us assume $j(\kappa)<\kappa$, and say that
$j(\kappa)=y$. Then since $\kappa$ is the smallest ordinal moved
by $j$ and $y<\kappa$ it must mean that $j(y)=y$. Because of
elementarity ordinal order is preserved, meaning that if
$y<\kappa$ then $j(y)<j(\kappa)$. But from the above we see that
$j(y)$ cannot be lesser than $j(\kappa)$ because
$j(y)=y=j(\kappa)$. And so we conclude that $j(\kappa)<\kappa$, is
not consistent.  Therefore it must be so that $j(\kappa)>\kappa$.

\section*{Acknowledgements}

I would like to thank Paul Corazza for his guidance as well as
B.J. Hiley, E. Sj\"{o}qvist, D. Abbott, C-G. Granqvist, P.
Nadel-Turo\'{n}ski, P. Hammerstein, E. Palmgren, J. L.
Garcia-Palacios, E.D. Avenda\~{n}o Soto, S. Mirbt, and D.
H\'{e}risson for discussions or comments.

\end{document}